\documentclass[conference]{IEEEtran}
\IEEEoverridecommandlockouts
\usepackage{cite}
\usepackage{amsmath,amssymb,amsfonts}
\usepackage{algorithmic}
\usepackage{graphicx}
\usepackage{textcomp}
\usepackage{xcolor}
\usepackage{balance}
\def\BibTeX{{\rm B\kern-.05em{\sc i\kern-.025em b}\kern-.08em
    T\kern-.1667em\lower.7ex\hbox{E}\kern-.125emX}}

\begin{document}


\title{Lost at Sea: Assessment and Evaluation of Rootkit Attacks on Shipboard Microgrids}

\author{
\IEEEauthorblockN{\textbf{Suman Rath}\IEEEauthorrefmark{1}, \textbf{Andres Intriago}\IEEEauthorrefmark{2}, \textbf{Shamik Sengupta}\IEEEauthorrefmark{1}, \textbf{Charalambos Konstantinou}\IEEEauthorrefmark{2}}
\IEEEauthorblockA{
\IEEEauthorblockA{\IEEEauthorrefmark{1}Department of Computer Science and Engineering, University of Nevada, Reno\\
\IEEEauthorrefmark{2}CEMSE Division, King Abdullah University of Science and Technology (KAUST)\\
}
E-mail: \{srath@nevada., ssengupta@\}unr.edu, \{andres.intriagovelasquez, charalambos.konstantinou\}@kaust.edu.sa}
}


\maketitle

\begin{abstract}
Increased dependence of the maritime industry on information and communication networks has made shipboard power systems vulnerable to stealthy cyber-attacks. One such attack variant, called rootkit, 
can leverage system knowledge to hide its presence and allow remotely located malware handlers to gain complete control of infected subsystems. This paper presents a comprehensive evaluation of the threat landscape imposed by such attack variants on Medium Voltage DC (MVDC) shipboard microgrids, including a discussion of their impact on the overall maritime sector in general, and provides several simulation results to demonstrate the same. It also analyzes and presents the actions of possible defense mechanisms, with specific emphasis on evasion, deception, and detection frameworks, that will help ship operators and maritime cybersecurity professionals protect their systems from such attacks.
\end{abstract}


\section{Introduction}
Electric ships play a significant role in the decarbonization of the maritime transport industry which supports more than 70\% of the total value of world trade including freight and passenger transportation \cite{rath2022microgrids}. Such systems are essentially microgrids with local power generation and consumption abilities and rely on interconnected digital systems to achieve control objectives like power sharing, propulsion, navigation, etc. This interconnected cyber layer is responsible for exchanging critical information that is in turn, responsible for maintaining the stability and reliability of such cyber-physical shipboard microgrids. As a consequence of this dependence, malicious attackers can manipulate cyber layer vulnerabilities (e.g., outdated software, insecure protocols, etc.) to lodge attack vectors and degrade system stability and control \cite{ospina2021evaluation, rath2022self}. Such attack vectors can also create equipment damage and power disruptions potentially jeopardizing human lives during maritime travel \cite{rath2022microgrids}. Several reports have highlighted that individual, large-scale cyber-attacks on the maritime sector can also disrupt the global supply chain and cause an estimated loss of more than \$110 billion USD \cite{daffron2019shen, bielby_2019}.

In order to address this problem, several researchers have focused on various cyber-attack types, operating principles, and proposed solutions for detection and mitigation \cite{kushal2018risk, zografopoulos2021cyber}. In this regard, malware-based infections (e.g., rootkits) have been studied by researchers as a particularly insidious threat against cyber-physical microgrids \cite{rath2022behind}. Rootkit refers to a highly lethal malware variant that can hide its presence inside infected hosts by using system knowledge to hide its actions \cite{krishnamurthy2019stealthy}. The rootkit is generally operated by a remotely located adversary who receives real-time system data collected by the malware via remotely accessible secret backdoors \cite{rath2021stealthy}. These backdoors can also be used by attackers to issue commands to the malware, execute targeted manipulation in host devices, and control the overall operation of the shipboard microgrid \cite{rath2022behind}.

Rootkit attacks on shipboard microgrids are a concerning threat that must be taken into account while designing embedded devices, security software, and cybersecurity operating policies about shipboard microgrids. Such attacks should also form an integral part of the vulnerability assessment and penetration testing procedures for industrial maritime equipment certification laboratories, especially concerned with reliability testing for power system devices. This paper seeks to extend our previous work on cybersecurity in shipboard power systems \cite{rath2022microgrids}, and investigate the impact of stealthy rootkit attacks executed via malware installed at one or more microgrid components. In the post-infection phase, the rootkit aggregates system knowledge in a stealthy manner to enable remote malware handlers to craft malicious attacks and fool possible defense systems. We consider two different stealthy attack models that adversaries can execute by leveraging system knowledge: \textit{(i)} drifting attacks where the rootkit injects false data (of gradually increasing magnitude) into critical signals and masks them as bias/noise, and \textit{(ii)} denial-of-service (DoS) attacks where the rootkit hibernates for its entire lifetime but activates itself to execute a DoS attack on the system once the ship is in a vulnerable location (e.g., in international waters or when a storm is approaching). Each of the presented attack models can have a devastating impact on the ship and severely jeopardize the lives of crew members, passengers, and cargo. The paper also provides an analysis of possible defenses to thwart such attacks from hardware, software, and network viewpoints emphasizing how such rootkit attacks can be evaded, deceived, detected, and mitigated. Simulation results are provided to showcase the possible impact of such attacks on shipboard microgrids and demonstrate the potential effect of mitigation mechanisms.

\section{Attack Model}

\subsection{Cyber-Physical Vulnerabilities in Shipboard Microgrids}
A typical shipboard microgrid consists of three control layers: \textit{(i)} the device control (DeCo) layer (also referred to as the primary controller), \textit{(ii)} the distributed power management (PM) layer (also referred to as the secondary controller), and \textit{(iii)} the Energy Management (EM) layer (also referred to as the tertiary controller) \cite{rath2022microgrids}. In this framework, the DeCo, the PM, and the EM layers are responsible for local generation control, bus-level voltage regulation, and load allocation, respectively. The PM layer is also responsible for ensuring that voltage levels adhere to a globally set reference setpoint and are synchronized for all the generators. This necessitates the use of a communication network and computing infrastructure which makes the layer vulnerable to malicious manipulations. Additional vulnerabilities also exist in shipboard microgrids as a consequence of sensors that can be spoofed and inadequately secured cyber/physical devices which can be subjected to unilateral manipulation for system state alteration.

Apart from inadequately secured sensors, other physical devices (e.g., embedded controllers, generators, actuators, etc.) can also be vulnerable to manipulation as a consequence of insider access and/or outsourcing from external manufacturers (supply chain). Physical devices are responsible for sending measurements to the cyber layer which computes control signal values. Hence, vulnerabilities in the physical layer can lead to a cascading effect that jeopardizes the operation of the entire cyber-physical shipboard microgrid. Cyber layer vulnerabilities in such systems are generally in the form of bad encryption, software bugs, expired security certificates, etc. Bad encryption allows attackers to decode data collected by the malware and devise stealthy manipulation strategies. Software loopholes allow the creation of backdoors for the extraction of critical state information. Further, expired security certificates impede the detection of high-volume attack traffic allowing remote attackers to stay hidden and continue manipulating the system for an elongated time period. Several real-world cyber-attacks on maritime industries and individual shipboards exploit one or more of the discussed security flaws to port malware. After malware installation, the attackers attempt to control the system. This was witnessed during the 2019 cyber-attack on a deep draft ship moving towards the port of New York where malware was found to be lodged in the system's control layer \cite{akpan2022cybersecurity, winder_2019}. Malware-based attacks may also attempt to steal/encrypt critical data (e.g., operational plans, employees' personal identification information, etc.) from shipping firms' servers paralyzing their operations and causing economic losses. Some of the recent examples of such attacks include the January 2023 attack on Norwegian firm DNV \cite{treloar_2023} and the March 2023 attack on Dutch logistics firm Royal Dirkzwager \cite{arghire_2023}. In several other examples, attackers demanded millions of dollars in ransom to release sensitive data and restore nominal operations for the breached firm.

\subsection{Rootkits: Characteristics and Operation}

\begin{figure}[t]
\centering
\centerline{{\includegraphics[width=0.9\linewidth]{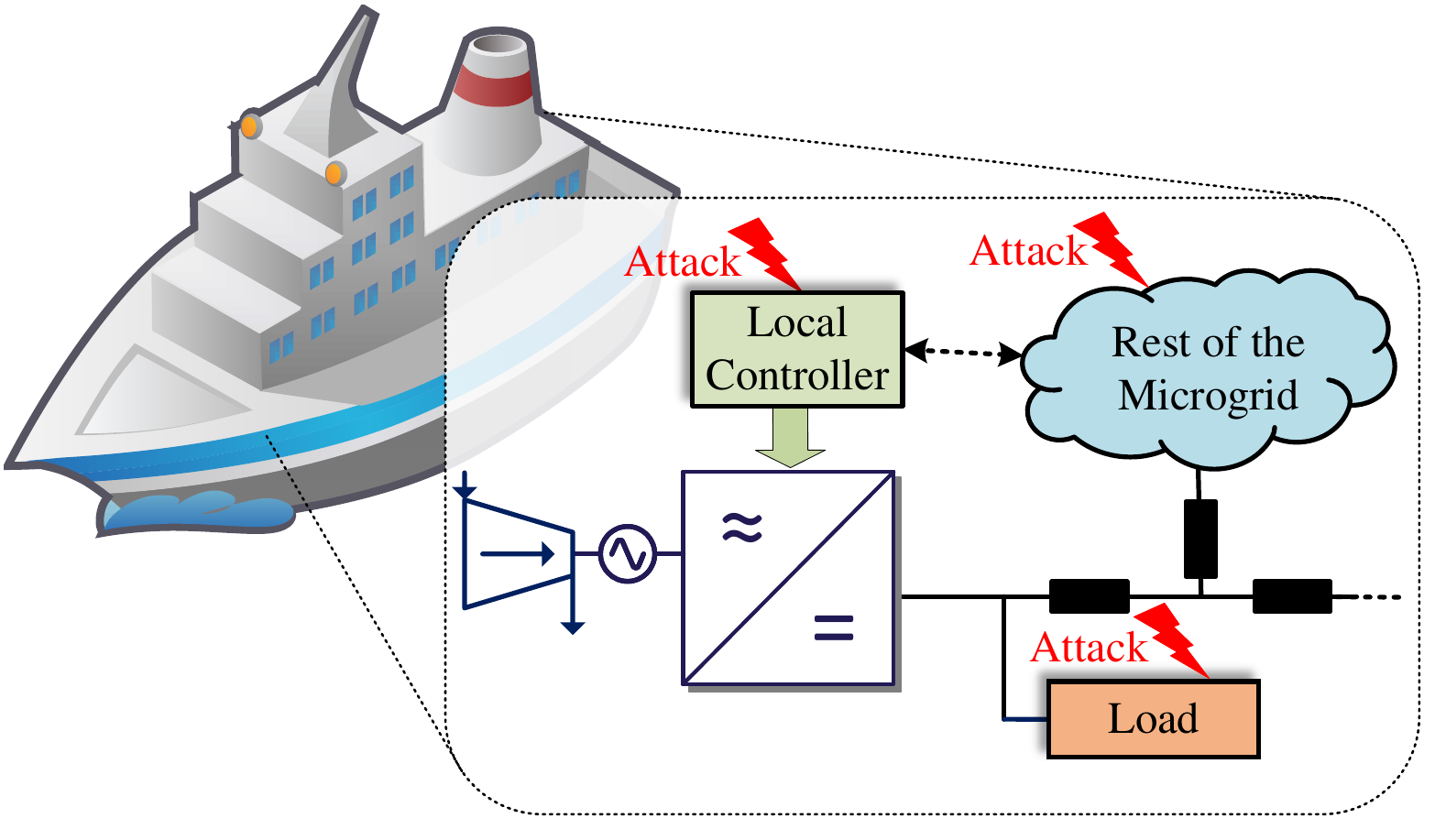}}}
\vspace{-3mm}
\caption{{Rootkit operation and possible locations in a shipboard microgrid.}}
\vspace{-4mm}
\label{fig:attack}
\end{figure}

Rootkits can generally penetrate electric shipboard systems via operating system/application layer vulnerabilities and get lodged at the kernel level inside its core computing systems to achieve privileged access. Once malware variants are installed, they seek to achieve persistence by action-masking via log modification and system knowledge exploitation \cite{rath2022behind}. In the conventional sense, the infected devices present the same vulnerabilities as general consumer-level devices meaning that rootkits can also access several hardware/software components including memory units, kernel, firmware, etc.  \cite{rath2021stealthy}. These vulnerabilities are often amplified in shipboard systems due to their isolated nature and lack of security protocols.

Fig. \ref{fig:attack} depicts the concept of malware locations on a shipboard microgrid. In this paper, it is assumed that the rootkit is installed at the measurement and control layers. It can eavesdrop and send information back to its handlers who then use the same to determine further course of action (target assignment and mode of operation). This course of action is generally scheduled in a manner that upholds the rootkit's stealthiness and persistence inside the host system(s). This is mainly achieved by ensuring that the manipulations performed are gradual and of low magnitude (e.g., minuscule bias/noise injections into individual sensors that flow into the control layer as inputs to produce erroneous decisions) to evade screening by bad data detectors. The high-level manipulations are extended over a long course of time to ensure attack propagation to other microgrid devices (e.g., controllers, power electronics, etc.).

In addition to high-level manipulations as depicted above, the rootkit may also create coordinated manipulations from multiple sensors to preserve inter-parameter relationships and hide low-level manipulations. The absence of preset reference voltage setpoints in shipboard microgrids (unlike terrestrial, grid-connected microgrid systems) means that the rootkit can also manipulate the local references to fool the system into tracking higher bus voltages leading to, e.g., component damage or lower bus voltages resulting in a DoS from critical subsystems (e.g., power generation modules (PGM), conversion-level devices, etc.) jeopardizing the overall ship operation. Rootkit attacks can also target load-sharing patterns among PGMs leading to load increases of one or more units, and thus, causing possible burnout of the module(s) and paralysis of shipboard power generation capabilities~\cite{nguyen2021advanced}. 

After the completion of an assigned target objective, the rootkit enters a passive mode where it eavesdrops on the infected system and continues to send data to its handlers until a new objective is assigned. This continues as a loop unless terminated by the remote malware handler(s). 



\section{Control Structure and Attack Formulation}
The shipboard microgrid is like any other autonomous DC microgrid \cite{batmani2022sdre}, but typically divided into two or more zones for robust operability and controllability \cite{rath2022microgrids}. Let the number of sources (with corresponding buses) in this system be $N$. Each of the sources in this microgrid is interconnected via AC-DC or DC-DC power converters which (along with sensory devices) form the physical layer of the system. As previously mentioned, voltage regulation and load current sharing (among the sources) is achieved in this system via the distributed PM layer \cite{rath2022microgrids, dav}. Each PGM in this system is associated with a corresponding local controller which attempts to exchange information about local measurements with neighboring nodes.

\begin{figure}[]
\centering
\centerline{{\includegraphics[width=0.92\linewidth]{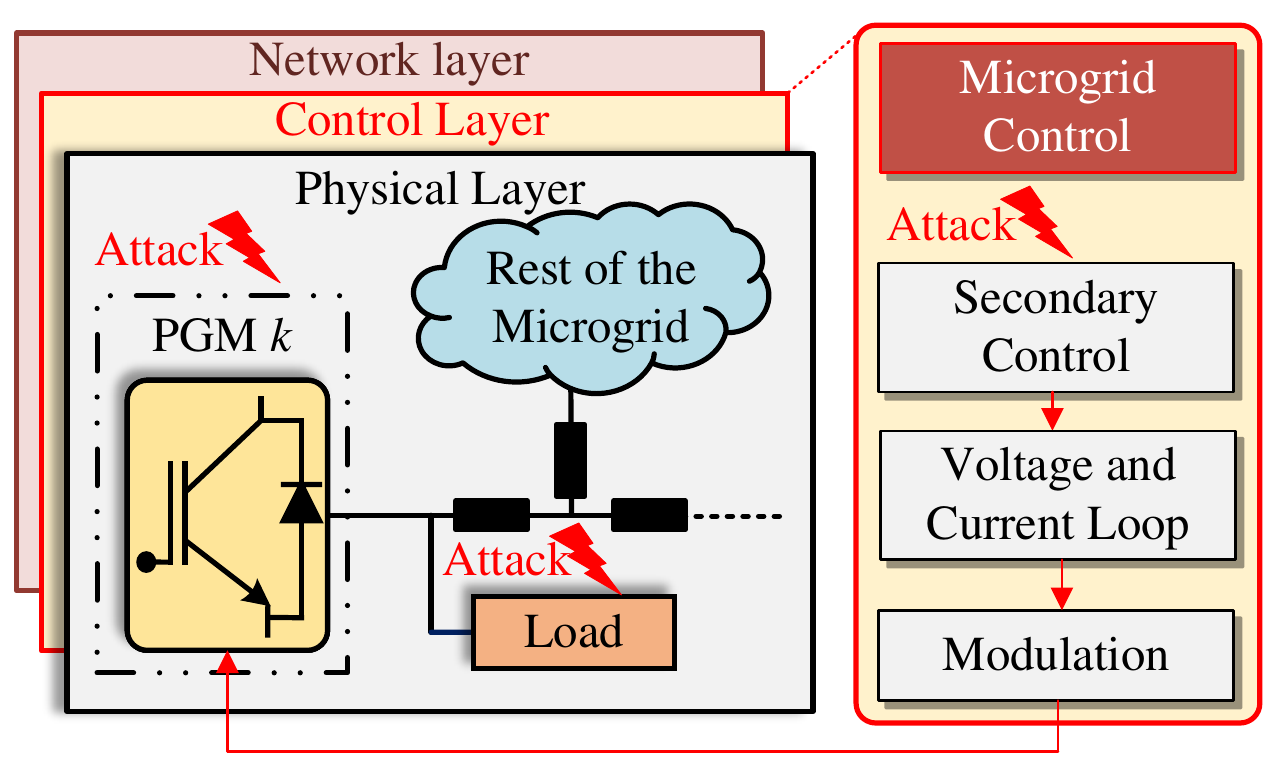}}}
\caption{{Schematic diagram of the control framework in shipboard microgrids and possible attack surfaces.}}
\label{fig:control}
\end{figure}

In the depicted system (Fig. \ref{fig:control}), the cyber layer is a graph with the PGM(s) (known hereafter as \textit{agents}) denoted by nodes and communication links represented by edges. The graph elements constitute a matrix ${A}=[a_{kj}] \in \mathbb{R}^{N \times N}$ with communication weights $a_{kj} > 0, \text{if} (\psi_{k}, \psi_{j}) \in {E}$ (${E}$ represents an edge and $\psi_k$ a node). For all other conditions, $a_{kj} = 0$. The inbound input matrix is defined as ${Z}_{\text{in}} = \texttt{diag}\{\sum_{k \in K} a_{kj}$\}. ${L}$ represents a Laplacian matrix which is considered to be balanced if ${A} = {Z}_{\text{in}}$ \cite{takiddin2022data}.

Each agent exchanges a series of local measurements with its neighboring nodes (for the attainment of voltage regulation and proportionate load-sharing goals) as shown below:
\begin{equation}
    {x} = \{\bar{{V}}, {I}^{\text{pu}}\}
\end{equation}
The control inputs to each local-level distributed secondary controller can be described as:
\begin{eqnarray}
    u_k(t) = \sum_{j \in N^E_k} \underbrace{a_{kj} (x_j(t) - x_k(t))}_{e_k(t)}
\label{1}
\end{eqnarray}
where $u_k$ is defined as \{$u^V_k, u^I_k$\} and $e_k$ as \{$e^V_k, e^I_k$\} (corresponding to elements in $x$), and $N^E_k$ represents neighboring nodes of $k$. The error term in (\ref{1}) can be simplified as:
\begin{eqnarray}
    e^V_{k}(t) = a_{kj}(\bar{V}_j(t) - \bar{V}_k(t))\\
    e^I_{k}(t) = a_{kj}({I}^{pu}_j(t) - {I}^{pu}_k(t))
\end{eqnarray}
$u_k$ can also be represented via extrapolation in a similar manner  \cite{takiddin2022data}.\\
\textbf{\textit{Remark:}} All distributed agents in the framework depicted above achieve consensus via a cooperative synchronization law \cite{sync} by enabling $\dot{{x}}(t) = -{L} {x}(t)$ to achieve convergence to $\lim\limits_{t \rightarrow \infty} x_k(t) = c, \ \forall \ k \in K$  \cite{takiddin2022data}.
    
The secondary controller provides the following inputs for the achievement of voltage- and load-sharing goals for the $k^{th}$ agent via the
usage of
voltage improvement equations \cite{sahoo2017distributed}:
\begin{eqnarray}
  \Delta V_{1_{k}} = H_1 (s) (V_{{ref}} - \bar{V}_{{k}}) \label{8}\\
  \Delta V_{2_{k}} = H_2 (s) {(I_{{ref}} - u^I_k)} \label{9}
\end{eqnarray} 
where $I_{{ref}}$ = 0 (for proportionate load current sharing) and $\bar{V}_{{k}}$ = $V_{{k}}$ + $\int_{0}^{\tau} \sum_{j \in N^E_k} u^V_k d\tau$.
Local-level voltage reference values for the $k^{th}$ agent can be obtained via the summation of the global reference voltage term and additional correction terms as obtained from (\ref{8}) and (\ref{9}) \cite{takiddin2022data}. This is depicted below:
\begin{equation}
    V^{k}_{{ref}} = V_{{ref}} + \Delta V_{1_{k}} + \Delta V_{2_{k}}. \label{10}
\end{equation}
Control targets formulated in (3) and (4) can be achieved using the reference voltage value depicted in (\ref{10}).
Considering an adequately connected digraph, the
microgrid system
objectives
would achieve convergence to:
    \begin{eqnarray}
    \lim\limits_{t \rightarrow \infty} \bar{V}_k (t) = V_{{ref}}
, \ \lim\limits_{t \rightarrow \infty} u^I_k(t) = 0 \ \ \forall \ k \ \in \ K. \label{11}    \end{eqnarray} 
		
		
As depicted in Figs. \ref{fig:attack} and \ref{fig:control}, the control structure of a shipboard microgrid is vulnerable to cyber-attacks that generally attempt to disrupt (\ref{11}). The rootkit attack hides in the kernel of multiple control devices to create low-magnitude deviations that persist over time to result in power outages which lead to control blackouts forcing the ship to have no decision-making power in the middle of a sea or in emergencies like approaching tornadoes. To prevent such incidents, it is essential that such malware variants are identified and removed.

Rootkits can also perform simultaneous, coordinated manipulations to hide their manipulations from system observers and bad data detectors. This can be depicted via the addition of an external term to (1):
\begin{eqnarray}
{u}^{a}(t) = {L} {x}(t) +  {W} {x}_{attack} \label{12}    \end{eqnarray}
where ${u}^a, {x}$, and ${x}_{attack}$ signify the vector form of input signal $u^a_k$ = \{$u^{Va}_k, u^{Ia}_k$\}, states as formulated by $x_k$ = \{$\bar{V}_{{k}}, I^{pu}_{{k}}$\}, and attack injections [$x^{V}_{attack_{k}}, x^{I}_{attack_{k}}]^T$, respectively. ${x}_{attack}$ may be bounded or unbounded depending on the target objective assigned to the rootkit. Further, $|x_{attack}| < |x^{Thres}_{attack}|$ where $|x^{Thres}_{attack}|$ is the threshold margin for incoming data to be classified as anomalous by the shipboard microgrid's generic bad data detectors. ${W}$ = [$w_{kj}$] is the rootkit access level matrix which is typically formulated as:
\begin{eqnarray}
w_{kj} = {\begin{cases}
	- \frac{1}{N^E_k + 1}, \ j \in N^E_k\\
	1 + \sum_{j \epsilon N^E_k} w_{kj}, \ j = k \\
	0, j \not\in N^E_k, j \neq k
	\end{cases}} \label{14}
\end{eqnarray}
In \eqref{14}, diagonal elements represent manipulations in local measurement signal ${x}$. Non-zero, off-diagonal terms in ${W}$ correspond to neighboring measurement values received via a communication network. A hidden attack model can be sustained via (9), if and only if the summation of state modifications and malware-induced state evolutions are a subset of the shipboard microgrid's weakly unobservable subspace. Considering a microgrid-level viewpoint, ${W}{x}_{attack} = 0$. However, the rootkit suppresses any identifiable change from one agent via the introduction of an oppositely directed shift corresponding to its neighboring measurements. This contributes a very minute change to the system dynamics of the shipboard microgrid making the manipulation unobservable at any randomly chosen instantaneous time slot.

\section{Defense Strategies against Rootkit Attacks}

This section provides a discussion of shipboard microgrid components (with specific notes on hardware, software, and network-layer components), their vulnerabilities, and possible defense strategies that can be employed to counteract rootkit attacks at each stage of the cyber kill chain.

The hardware components in shipboard microgrids generally consist of programmable logic controllers (PLCs), inverters, sensors, actuators, and energy storage devices (ESDs) \cite{song2020recursive}. Each of these components may be vulnerable to rootkit attacks trying to create a deviation in the state trajectory to achieve specific objectives. On the hardware side, rootkits may try to manipulate the firmware or basic input/output system (BIOS)-level code to achieve such objectives \cite{rath2022behind, rath2021stealthy}. Rootkit malware variants can also modify low-level code responsible for hardware control to manipulate overall system dynamics and bypass bad data detectors. Hence, significant improvements are needed to increase the resiliency of hardware components against rootkit attacks. Some of the possible defenses against such attacks are secure boot, hardware attestation, firmware integrity checks, etc. \cite{kiperberg2015remote, seshadri2005pioneer, wang2015confirm}. Secure boot ensures that only authorized software and/or trusted code can be loaded during the system startup so as to prevent malicious agents  from gaining access \cite{wen2010implicit}. Regular firmware integrity checks can also be utilized to detect any changes to the hardware firmware or BIOS code and alert microgrid operators about potential security breaches. Attestation strategies (e.g., \cite{kiperberg2015remote}) can protect microgrid hardware against rootkits via integrity authentication to identify tampering attempts \cite{seshadri2005pioneer}.

Hardware-based security modules (HSMs) can also be used to protect shipboard microgrid hardware against rootkits. These modules provide a secure way to store digital cryptographic keys and other sensitive data to protect them from eavesdropping and tampering \cite{seol2015trusted}. Ship defenders can utilize the unavailability of knowledge about encryption to ensure that any possible state information that is stolen by the malware is rendered undecipherable to its handlers. Additionally, intrusion detection and prevention systems (IDS/IPS) can also be used to protect microgrid hardware \cite{mclaughlin2016cybersecurity}. These systems can be configured to identify signature-based/behavioral analytics and alert system defenders about a potential breach.

Software components in shipboard microgrids typically consist of human-machine interfaces (HMIs), microgrid control and energy management systems. Each of these components may present several vulnerabilities that the rootkit can use to manipulate. Some of the possible software vulnerabilities are code errors, insecure software updates, and improper access control measures (e.g., weak passwords, unsecured communication links, network connections, etc.) \cite{rath2022microgrids, rath2022behind, rath2020cyber}. Rootkits can often hide their presence in low, kernel-level components which makes their detection very hard. They can also use their access level to modify the software or inject malicious code into the system, compromising its security and integrity. Hence, defenses must be developed to protect critical shipboard software against such malware-induced attack variants. Some of the possible defenses include regular code reviews via manual/deep learning agents for vulnerability identification and analysis. Code reviews could be made mandatory after software updates to find malicious loopholes.  Moreover, such updates could also be verified via digital signatures to prevent untrusted parties from gaining access to the system \cite{rath2022self, rath2020cyber}. Besides the generic security measures discussed above, runtime integrity checks \cite{feng2018behaviorki}, software-level IDS/IPS \cite{ali2020novel}, and appropriate access controls \cite{jha2008towards} can also be employed to ensure the prevention, detection, and mitigation of rootkit attacks. Access controls (e.g., two-factor authentication, strong passwords, etc.) are a vital component in this aspect as they can prevent the entry of malicious handlers to hinder rootkit deployment into the system.

Apart from the hardware and software vulnerabilities as depicted in the preceding paragraphs, shipboard microgrids may also contain several vulnerabilities at the network layer (e.g., insecure communication channels/network protocols). At the network level, rootkits can eavesdrop on system information and modify network traffic as well, compromising microgrid security and privacy. Moreover, these vulnerabilities can also be used to manipulate sensor measurements that are flowing to individual controllers to generate erroneous control decisions leading to system instability \cite{rath2022behind, takiddin2022data}. To eliminate these vulnerabilities, several network-based defenses can be utilized including encryption, network segmentation, honeypots, moving target defenses (MTD), access control, etc. \cite{spitzner2003honeypots, rath2022behind, rath2020cyber, lakshminarayana2019moving}. Encryption and digital signatures can be used to make sure that no stolen information may be used to manipulate system dynamics \cite{rath2020cyber}. On a similar note, network segmentation and honeypots can be used to elevate protection measures for critical subsystems (e.g., PGMs). Network segmentation can isolate sensitive system components from open-access subsystems like infotainment units. Similarly, honeypots can be used to detect the presence of possible attackers and lure them away from critical microgrid components. MTD can be utilized to make dynamic changes to the microgrid configuration adding an uncertainty factor and decreasing the success probability of attacks. Access control can be used to prevent unauthorized access to the microgrid network potentially preventing masking of manipulations by the rootkit handlers.

Clearly, a combination of hardware, software, and network-based defenses is essential to protect shipboard microgrids against rootkit attacks. Each of these measures can be used to thwart the actions of the malware handlers at one or more stages of the rootkit cyber kill chain \cite{rath2022behind}. However, apart from the technical defenses, human factors also play an important role in ensuring cyber-resiliency against rootkit attacks \cite{pugliaresi2013us}. Human factors can follow a series of best practices to thwart the installation of rootkits. It is essential that electric ship personnel are aware of cybersecurity policies and best practices, and are trained to implement security procedures in cases of emergency \cite{tam2018maritime}. This can be achieved by conducting regular training and exercises to test the effectiveness of the shipboard attack-response procedures. It is of paramount importance that crew members are trained and encouraged to report potential cyber threats and/or incidents. Whistle-blowers should be incentivized by the management and a culture of cyber-awareness (irrespective of designation/occupation) should be promoted and integrated into all aspects of the shipboard operations. 
In general, it is essential to ensure that a comprehensive, proactive approach is adopted to protect shipboard microgrids against rootkit attacks. A combination of technical defenses, best practices, and a strong cybersecurity culture is very important to provide a safe and reliable operational framework that protects the security and integrity of the system and ensures the safety of those on board.

\section{Simulation and Results}

\begin{figure}[]
\centering
\centerline{{\includegraphics[width=0.95\linewidth]{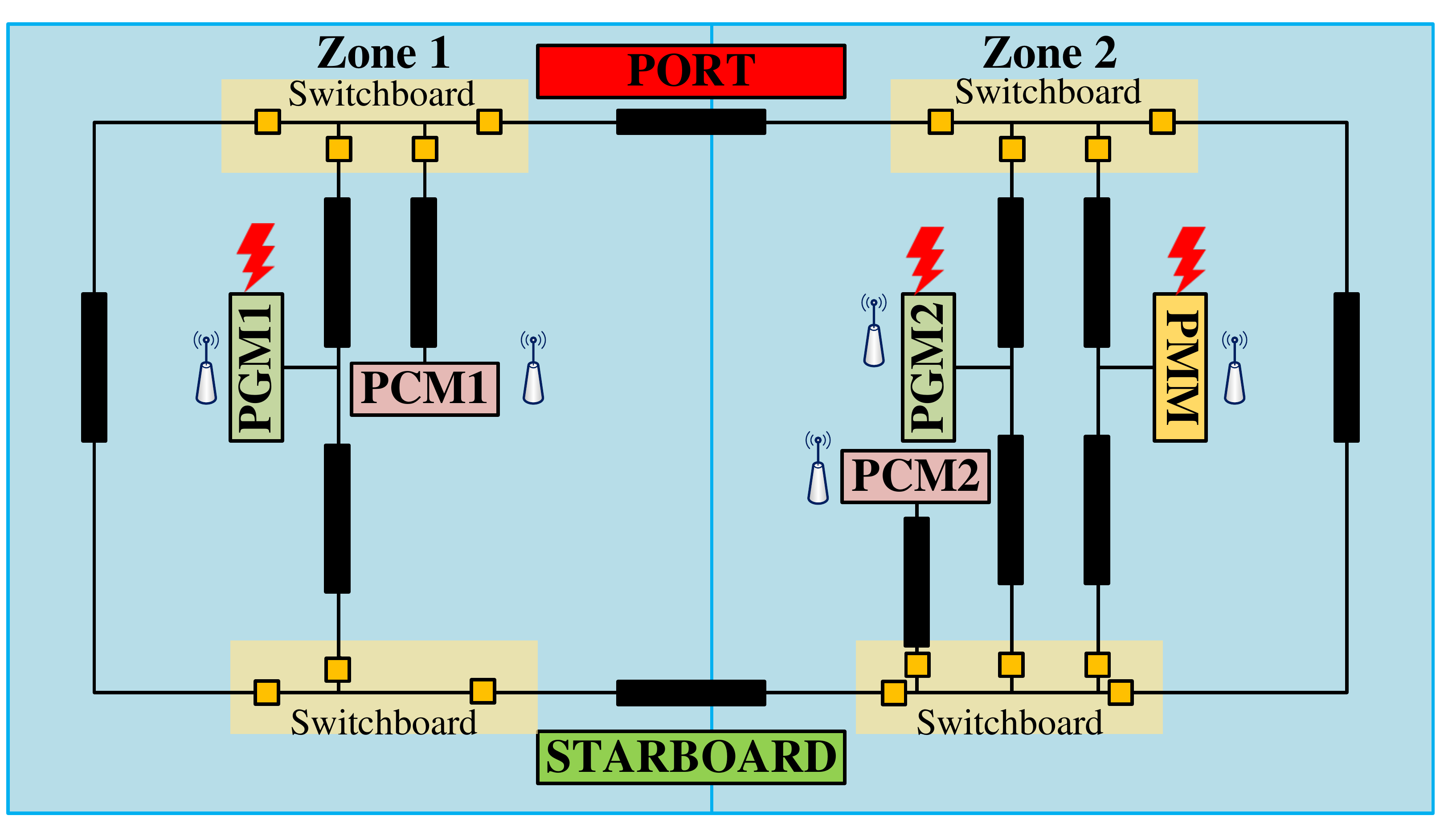}}}
\caption{{Schematic diagram of the dual-zone shipboard microgrid test system designed in the MATLAB environment \cite{rath2022microgrids}.}}
\label{fig:case}
\end{figure}

A dual-zone, notional MVDC shipboard microgrid test system is designed in MATLAB as detailed in \cite{ravindra2016documentation}. The schematic depiction of this test model is shown in Fig. \ref{fig:case}. Each zone in the system consists of a PGM and a power conversion module (PCM). Power sharing among the PGMs is regulated through voltage droop control. In addition to the PGM and the PCM, the second zone also consists of a propulsion motor module (PMM). Both zone 1 and zone 2 are connected to each other through the starboard and port side via switches. All subsystems are connected to one or more other subsystems via cables. Each PGM in the system consists of three different components: an AC/DC converter (for rectification before the generated power flows into the MVDC distribution system), an output filter, and the generator (which is a gas turbine of 36 MW capacity, i.e., the source). In the normal setting,
the distribution system is maintained at a constant DC voltage of 12 kV. The PCMs are rated at 10 MW each and are typically utilized for power conversion to lower voltages \cite{rath2022microgrids, xie2017comparative}. The PMM that is only present in the second zone, consists of two sub-units rated at 18 MW each and are of constant impedance with resistive loads. The switches allow this module to be fed either from the starboard or the port side. Detailed descriptions of the system parameters and their magnitudes are available in \cite{ravindra2016documentation}. Data collected from this system is analyzed via a deep learning framework as explained in \cite{rath2022behind, rath2021stealthy} to design stealthy rootkit attacks that can manipulate the system at the process level. We study the impact of rootkit attacks on this test system via several case studies as depicted in the following subsections. We mainly demonstrate how the rootkit can create drifting manipulations where it continues to inject false data into the original stream of \texttt{True} measurement inputs and how it can cause major deviations to cause system shutdown when the ship is at a vulnerable position (e.g., navigating in international waters during storms). Further, we also demonstrate how such manipulations can be detected using artificial intelligence (AI)-enabled behavioral analytics for timely mitigation.

\subsection{Target: Drifty Manipulation of System Voltage Levels}
\begin{figure}[]
\centering
\centerline{{\includegraphics[width=0.8\linewidth]{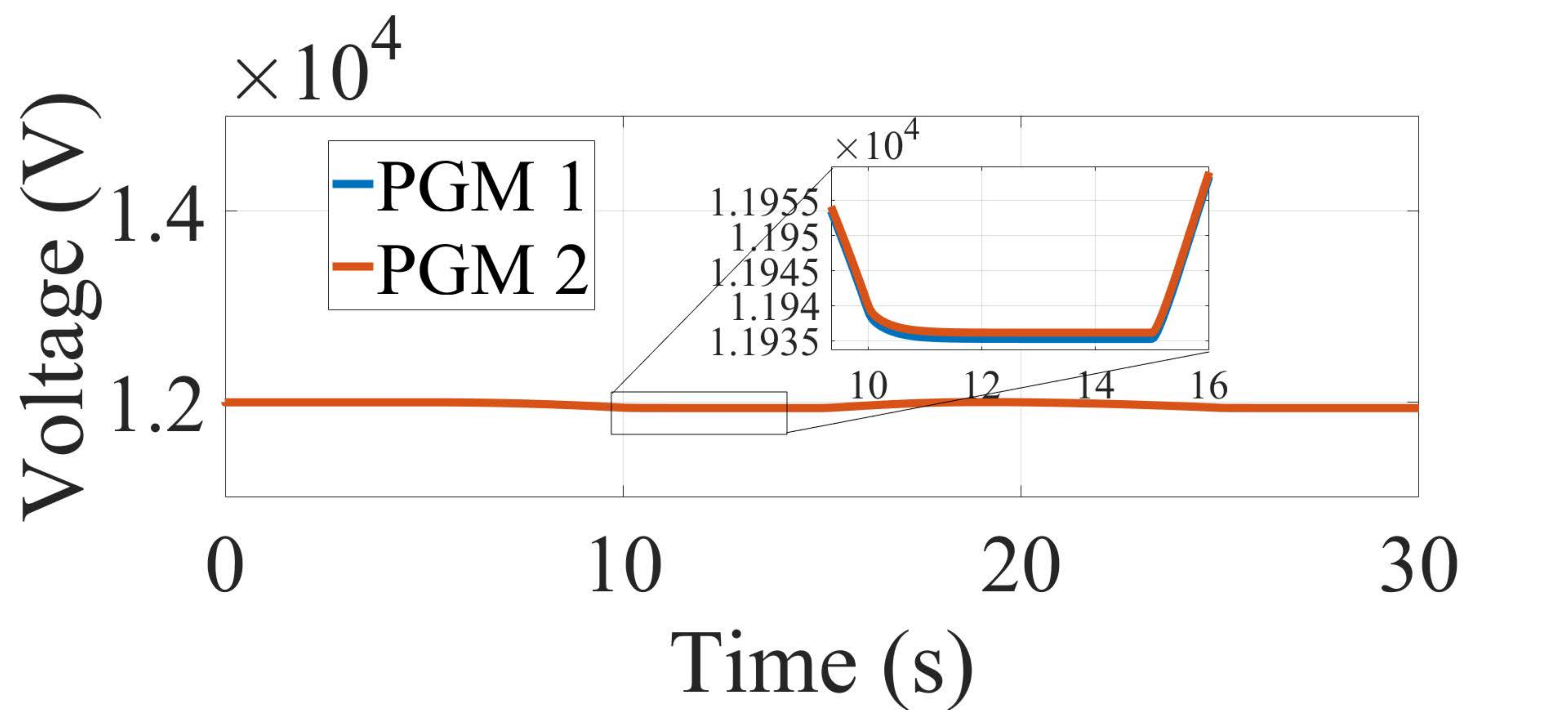}}}
\caption{{Voltage at PGM 1 and PGM 2 in the nominal scenario.}}
\label{fig:res1}
\end{figure}

\begin{figure}[]
\centering
\centerline{{\includegraphics[width=0.8\linewidth]{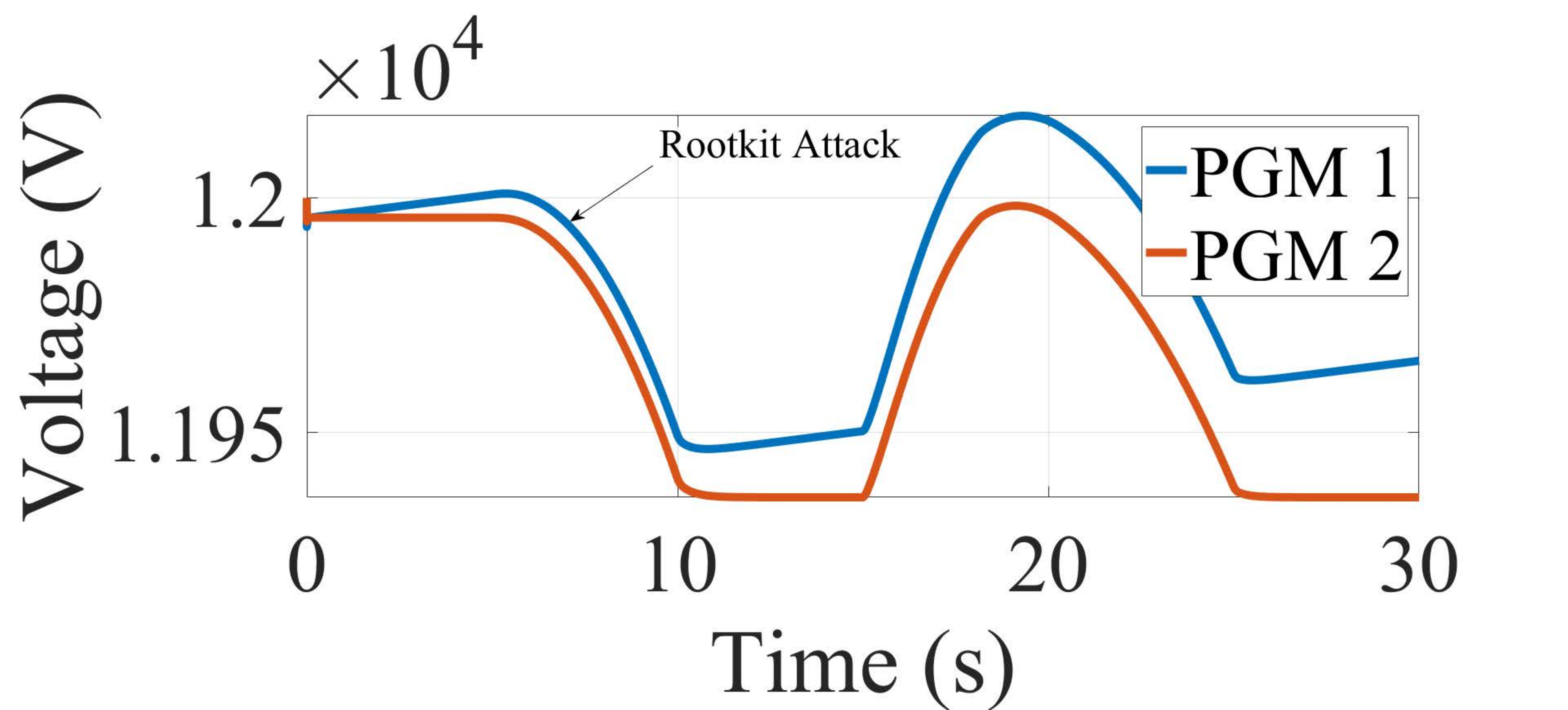}}}
\caption{{Magnitude of the system voltages (measured at PGM 1 and PGM 2) during the rootkit attack at PGM 1.}}
\label{fig:res2}
\end{figure}
When the remotely located malware handler instructs the rootkit to initiate a drifting attack on system levels, it analyzes the minimum number of infected host devices it can manipulate to achieve the desired deviation level. For achieving a higher voltage level at the handler-assigned locations, the rootkit typically needs to create manipulations at the measurement and control layers. The rootkit alters $\bar V_k$ to manipulate the error term in (3).
Fig. \ref{fig:res1} and Fig. \ref{fig:res2} show the system voltage levels in the nominal and manipulated scenarios, respectively. The rootkit successfully increases the output voltage of PGM 1 in an asynchronous manner to create system instability. Further, the uncontrolled drift-like increase of output voltages can damage end devices (e.g., consumer loads, propulsion motors, etc.) that receive power directly from PGM 1. This can serious repercussions if the ship is at a vulnerable location where it does not have access to alternate sources of power.

\subsection{Target: Disruptive Load Current Increase on PGM 1}
\begin{figure}[]
\centering
\centerline{{\includegraphics[width=0.8\linewidth]{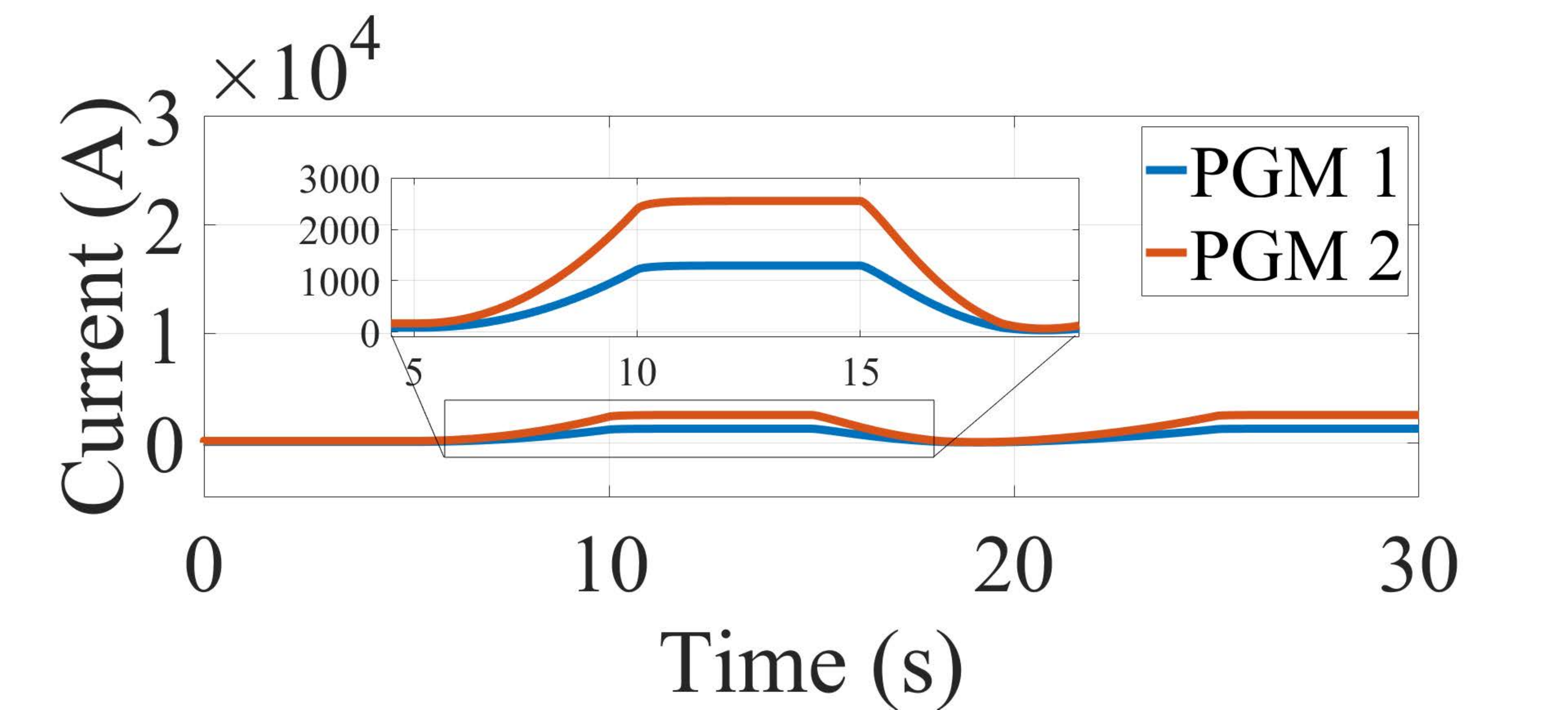}}}
\caption{{Load current at PGM 1 and PGM 2 in the nominal scenario.}}
\label{fig:res3}
\end{figure}

\begin{figure}[]
\centering
\centerline{{\includegraphics[width=0.8\linewidth]{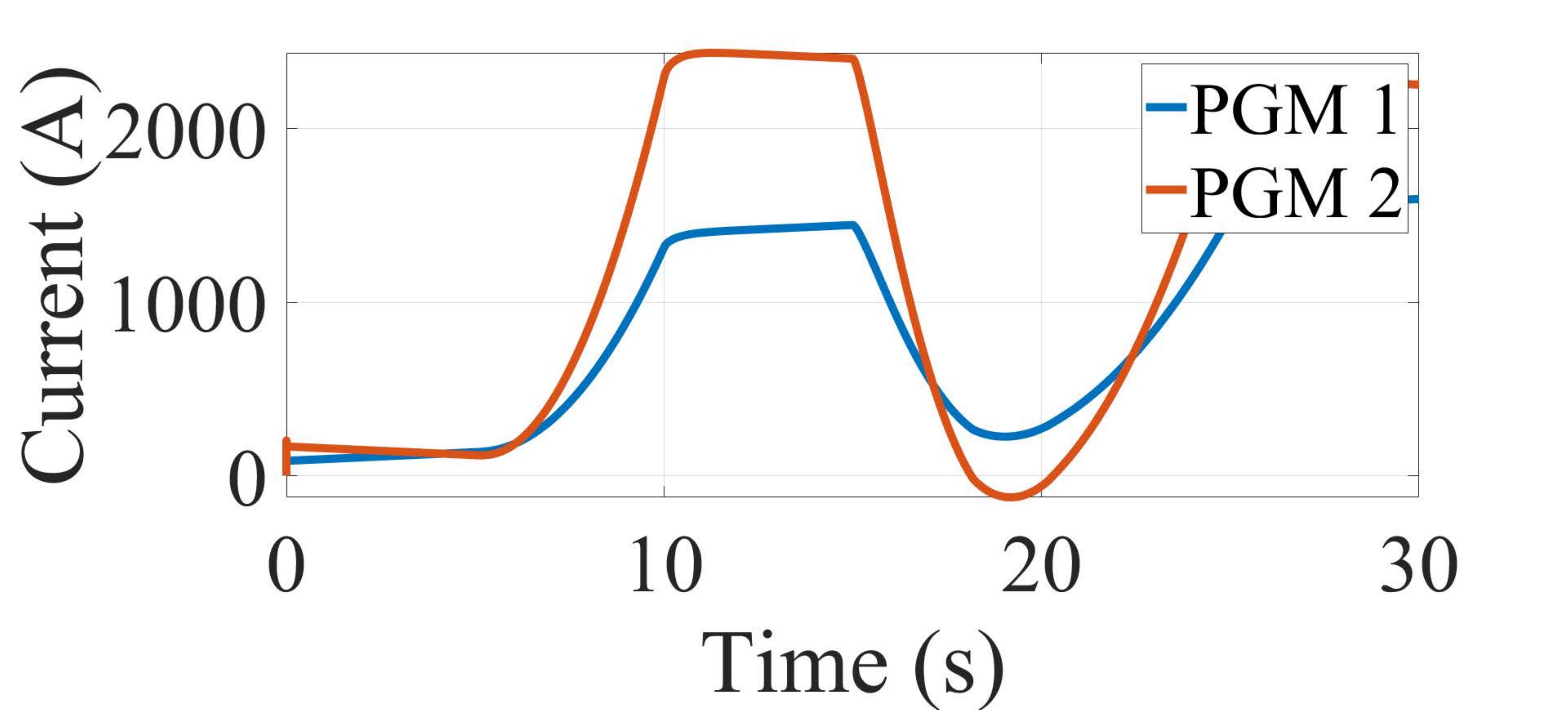}}}
\caption{{Load current at PGM 1 and PGM 2 during the rootkit attack.}}
\label{fig:res4}
\end{figure}

The test system depicted in Fig. \ref{fig:case} achieves an optimally configured load current-sharing arrangement among PGM 1 and 2 in the nominal, attack-free setting. However, if the attacker instructs the rootkit to disrupt this optimal load-sharing plan, the rootkit scans its access level to identify the smallest subset of devices which it can manipulate to achieve the assigned target. Another major factor in the identification of the target subset is the potential attack time frame which the attacker specifies along with the target assignment. Fig. \ref{fig:res3} and Fig. \ref{fig:res4} depict the load current at PGM 1 and PGM 2 during the normal and manipulated scenarios, respectively. Persistent manipulations may result in component shutdown and system blackouts after a gradual course of time.

\subsection{Target: System Shutdown}
\begin{figure}[t]
\centering
\centerline{{\includegraphics[width=0.8\linewidth]{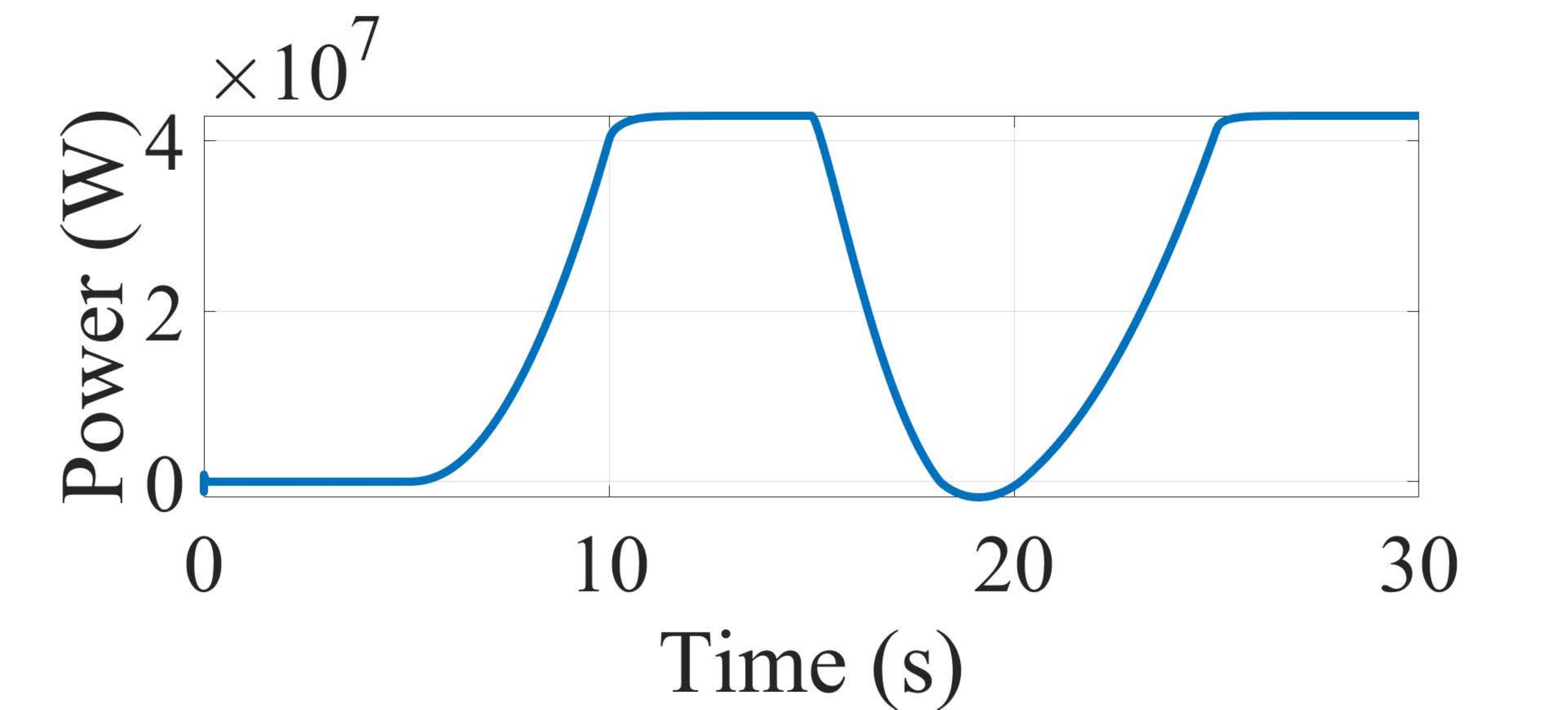}}}
\caption{{Power magnitude as measured at the PMM in the nominal scenario.}}
\label{fig:res5}
\end{figure}

\begin{figure}[t]
\centering
\centerline{{\includegraphics[width=0.8\linewidth]{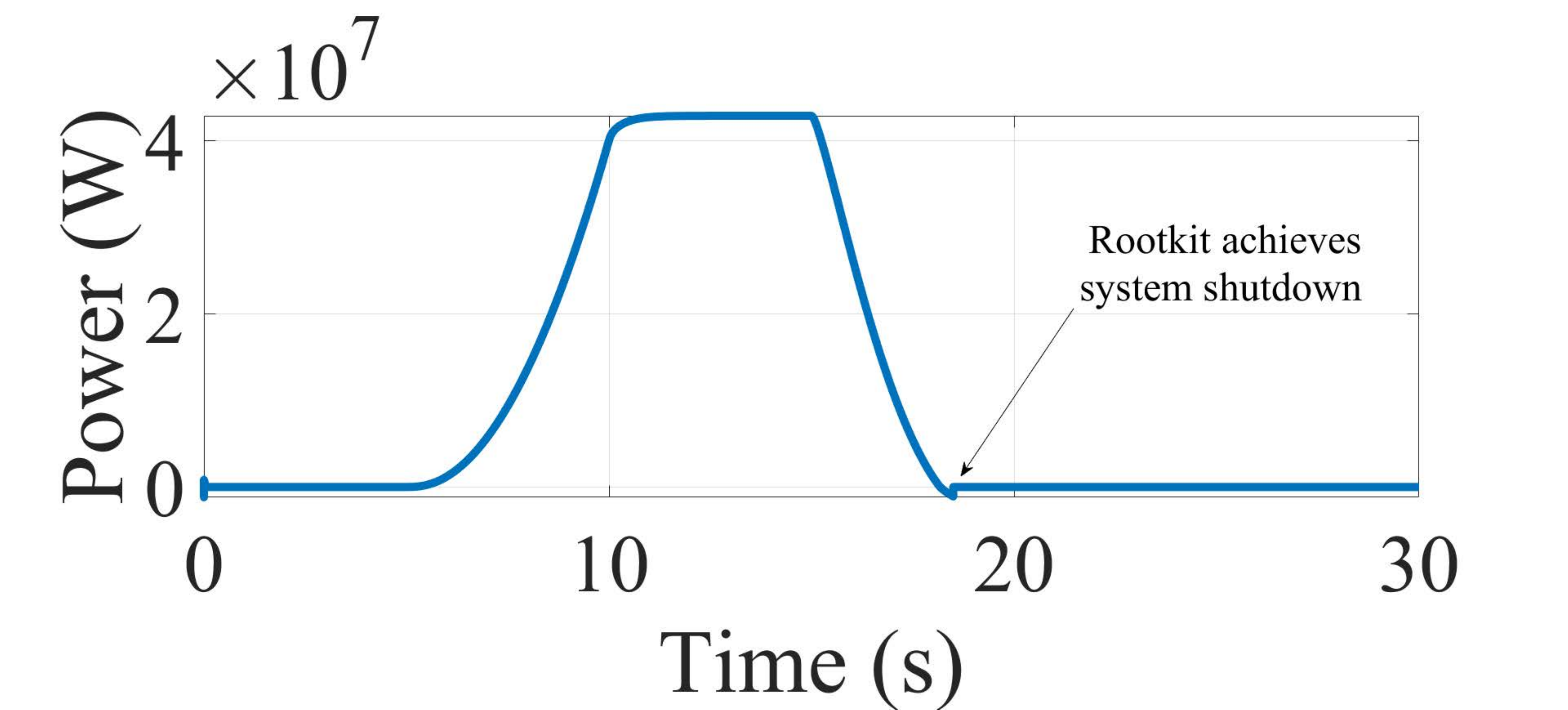}}}
\caption{{Power magnitude as measured at the PMM during a system shutdown caused by the rootkit attack.}}
\label{fig:res6}
\end{figure}
A shutdown of the system essentially means that the critical subsystems of the shipboard are turned off leading it into an uncontrollable trajectory. If the rootkit is requested to ``create'' a shutdown, it generally aims to perform this action when the ship does not have access to an alternate source of power that can allow it to recuperate. A typical example of a vulnerable time period would be when the ship is trying to navigate through international waters during heavy storms. Let $t = 18.5 s$ be a vulnerable time slot as identified by the rootkit handlers. Fig. \ref{fig:res5} demonstrates the magnitude of power measured at the ship's propulsion motor during the normal scenario. Fig. \ref{fig:res6} presents the power magnitude during the action of the rootkit that leads to system shutdown during the time slot initiated at $t = 18.5 s$. 

\subsection{Assessment: Behavioral Analytics for Rootkit Detection}
As depicted in Section IV, there are numerous ways in which ship operators and cybersecurity professionals may be able to counter rootkit attacks. One such strategy is the use of AI-enabled behavioral analytics for the detection of rootkit manipulations and possible mitigation. These strategies generally identify and extract features of nominal system data that are assumed to be free from attack elements. Such data elements can be generated by allowing the system to execute a \textit{dry run} when it is docked at a secure shipyard.
The generated data is fed into an artificial neural network (ANN)-based predictor model with two hidden layers to perform supervised training and enable fine-tuning of model parameters in terms of weights and bias magnitudes. The training procedure is initiated via a randomly chosen weight vector $W_r = [w_r]$. The role of the ANN is to generate a predicted signal value $x_p$, which is used to detect if an incoming signal has been manipulated. During the $i^{th}$ iteration of training, the predicted output $x_{p_i}$ is formulated as: 
\begin{equation}
    x_{p_i} = f_{ac}(w_{r_i}x_{o_i}+\zeta_i, l)
\end{equation}
where $f_{ac}(\circ)$ represents the activation function, $l$ represents the total number of layers, $x_o$ represents the \texttt{True} data element as obtained from the training dataset, and $w_{r_i}$ and $\zeta_i$ represent the weight
and bias
values during the $i^{th}$ iteration, respectively. The mean squared error ($MSE$) between $x_{p}$ and $x_o$ is computed as follows:
\begin{equation}
    MSE = \frac{1}{S}\sum_{i=1}^{S}(x_{pi}-x_{oi})^2
\end{equation}
where $S$ is the number of samples. Post-computation of $MSE$, the model parameters are updated (via backpropagation as depicted in \cite{rath2022behind}) and training is repeated again. This process continues until $MSE$ converges to a desired magnitude as specified by the defender. The final model parameters are denoted as $\{w_f, \zeta_f\}$. After real-time deployment, the model is used to generate $x_p(t)$ which is the predicted signal magnitude at a specified time-step $t$.

\begin{figure}[]
\centering
\centerline{{\includegraphics[width=0.5\linewidth]{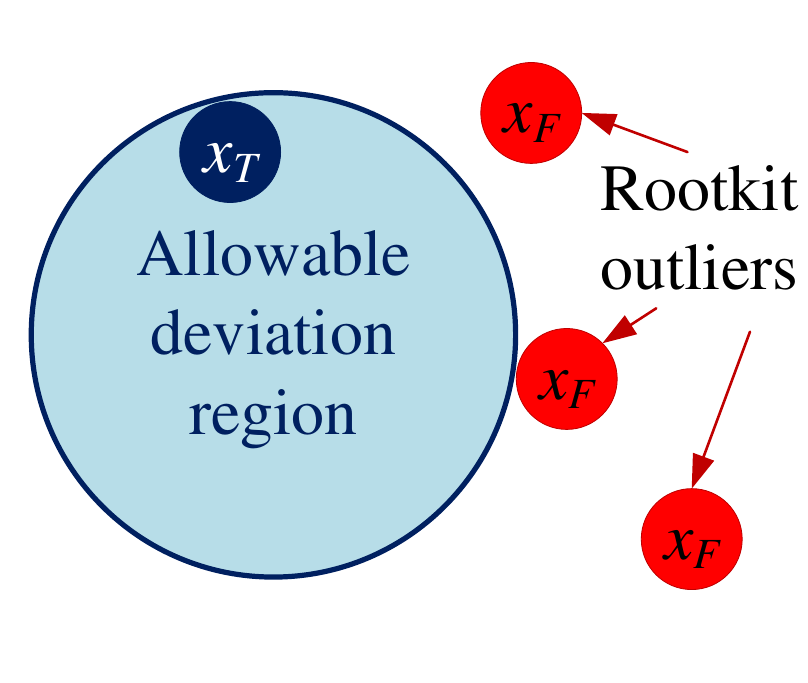}}}
\caption{{AI-enabled behavioral concept for the detection of rootkit manipulations.}}
\label{fig:threshold}
\end{figure}

\begin{figure}[]
\centering
\centerline{{\includegraphics[width=0.8\linewidth]{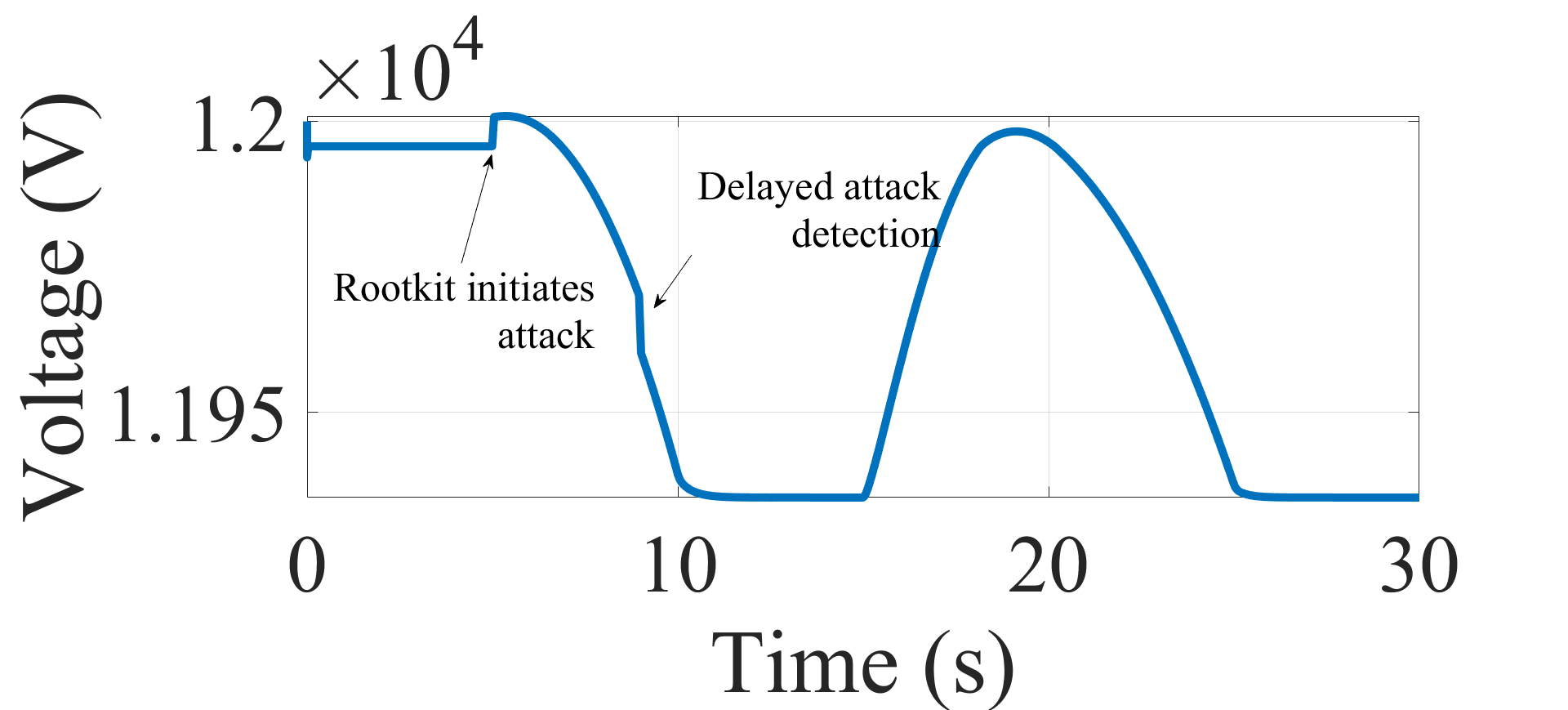}}}
\caption{{Performance of AI-enabled behavioral analytics in detecting and mitigating rootkit attacks.}}
\label{fig:res7}
\end{figure}

Further, $x_p(t)$ is used to create an \textit{allowable deviation region} which is a circular region centered at $x_p(t)$ with an infinitesimally small radius $\rho$ as depicted in Fig. \ref{fig:threshold}. The circular region represents the range in which incoming data points must lie in order for the system to remain stable. If the system encounters a data element lying outside this margin it is classified as an \textit{anomaly} and rejected by the recipient. Further, the rejected signal is reconstructed using the ANN model and fed into the recipient controller allowing it to function in a normal manner. This is depicted in Fig. \ref{fig:res7}. The system can identify outliers injected by the rootkit and replace them with \textit{reconstructed} values of the signal allowing it to function normally in a rootkit-infected environment.
A limitation of using behavioral analytics for rootkit detection is its inability to identify rootkit manipulations that lie within the \textit{allowable deviation region}. Hence, it needs significant improvement for deployment against process-aware rootkit attacks.


\section{Conclusion}
Rootkits are a major threat to shipboard microgrids as they can eavesdrop on system data and use it to mask their manipulations. This paper assesses the possible impact of rootkit attacks on such systems and demonstrates how they can perform stealthy manipulation of signals in critical subsystems. We showcase how the rootkit can create drifting attacks on system parameters including voltage and load current magnitudes. Further, we also provide a detailed discussion of how rootkit attacks can be countered at the hardware, software, and network-level components of shipboard microgrids. Finally, we present the performance of one of the discussed strategies, AI-enabled behavioral analytics via simulation results and explain its consequences and limitations. 

\balance
\bibliographystyle{IEEEtran}
\bibliography{biblio}

\begin{thebibliography}{10}
\providecommand{\url}[1]{#1}
\csname url@samestyle\endcsname
\providecommand{\newblock}{\relax}
\providecommand{\bibinfo}[2]{#2}
\providecommand{\BIBentrySTDinterwordspacing}{\spaceskip=0pt\relax}
\providecommand{\BIBentryALTinterwordstretchfactor}{4}
\providecommand{\BIBentryALTinterwordspacing}{\spaceskip=\fontdimen2\font plus
\BIBentryALTinterwordstretchfactor\fontdimen3\font minus
  \fontdimen4\font\relax}
\providecommand{\BIBforeignlanguage}[2]{{%
\expandafter\ifx\csname l@#1\endcsname\relax
\typeout{** WARNING: IEEEtran.bst: No hyphenation pattern has been}%
\typeout{** loaded for the language `#1'. Using the pattern for}%
\typeout{** the default language instead.}%
\else
\language=\csname l@#1\endcsname
\fi
#2}}
\providecommand{\BIBdecl}{\relax}
\BIBdecl

\bibitem{rath2022microgrids}
S.~Rath, C.~Konstantinou, B.~Papari, C.~S. Edrington, P.~Ge, and F.~Teng,
  ``Microgrids in mission-critical applications.''\hskip 1em plus 0.5em minus
  0.4em\relax Institution of Engineering and Technology, 2022.

\bibitem{ospina2021evaluation}
J.~Ospina, C.~Konstantinou, M.~Stanovich, and M.~Steurer, ``Evaluation of
  communication network models for shipboard power systems,'' in \emph{2021
  IEEE Electric Ship Technologies Symposium (ESTS)}.\hskip 1em plus 0.5em minus
  0.4em\relax IEEE, 2021, pp. 1--9.

\bibitem{rath2022self}
S.~Rath, L.~D. Nguyen, S.~Sahoo, and P.~Popovski, ``Self-healing secure
  blockchain framework in microgrids,'' \emph{IEEE Transactions on Smart Grid
  (\textit{Early Access})}, 2023.

\bibitem{daffron2019shen}
J.~Daffron, S.~Ruffle, A.~Coburn, J.~Copic, K.~Quantrill, K.~Strong, and
  E.~Leverett, ``Shen attack: Cyber risk in asia pacific ports,'' \emph{Centre
  for Risk Studies, Cambridge}, 2019.

\bibitem{bielby_2019}
\BIBentryALTinterwordspacing
K.~Bielby, ``A maritime cyber attack could cost $\$$110 billion and cripple
  global supply chains,'' Oct 2019. [Online]. Available:
  \url{https://www.hstoday.us/subject-matter-areas/transportation/a-maritime-cyber-attack-could-cost-110-billion-and-cripple-global-supply-chains/}
\BIBentrySTDinterwordspacing

\bibitem{kushal2018risk}
T.~R.~B. Kushal, K.~Lai, and M.~S. Illindala, ``Risk-based mitigation of load
  curtailment cyber attack using intelligent agents in a shipboard power
  system,'' \emph{IEEE Transactions on Smart Grid}, vol.~10, no.~5, pp.
  4741--4750, 2018.

\bibitem{zografopoulos2021cyber}
I.~Zografopoulos, J.~Ospina, X.~Liu, and C.~Konstantinou, ``Cyber-physical
  energy systems security: Threat modeling, risk assessment, resources,
  metrics, and case studies,'' \emph{IEEE Access}, vol.~9, pp.
  29\,775--29\,818, 2021.

\bibitem{rath2022behind}
S.~Rath, I.~Zografopoulos, P.~P. Vergara, V.~C. Nikolaidis, and
  C.~Konstantinou, ``Behind closed doors: Process-level rootkit attacks in
  cyber-physical microgrid systems,'' in \emph{2022 IEEE Power \& Energy
  Society General Meeting (PESGM)}.\hskip 1em plus 0.5em minus 0.4em\relax
  IEEE, 2022, pp. 1--10.

\bibitem{krishnamurthy2019stealthy}
P.~Krishnamurthy, H.~Salehghaffari, S.~Duraisamy, R.~Karri, and F.~Khorrami,
  ``Stealthy rootkits in smart grid controllers,'' in \emph{2019 IEEE 37th
  Int'l Conference on Computer Design (ICCD)}.\hskip 1em plus 0.5em minus
  0.4em\relax IEEE, 2019, pp. 20--28.

\bibitem{rath2021stealthy}
S.~Rath, I.~Zografopoulos, and C.~Konstantinou, ``Stealthy rootkit attacks on
  cyber-physical microgrids: Poster,'' in \emph{Proceedings of the Twelfth ACM
  Int'l Conference on Future Energy Systems}, 2021, pp. 294--295.

\bibitem{akpan2022cybersecurity}
F.~Akpan, G.~Bendiab, S.~Shiaeles, S.~Karamperidis, and M.~Michaloliakos,
  ``Cybersecurity challenges in the maritime sector,'' \emph{Network}, vol.~2,
  no.~1, pp. 123--138, 2022.

\bibitem{winder_2019}
\BIBentryALTinterwordspacing
D.~Winder, ``U.s. coast guard issues alert after ship heading into port of new
  york hit by cyberattack,'' Jul 2019. [Online]. Available:
  \url{https://www.forbes.com/sites/daveywinder/2019/07/09/u-s-coast-guard-issues-alert-after-ship-heading-into-port-of-new-york-hit-by-cyberattack/?sh=1135da7441aa}
\BIBentrySTDinterwordspacing

\bibitem{treloar_2023}
\BIBentryALTinterwordspacing
S.~Treloar, ``Cyber attack hits 1,000 merchant ships as norway firm targeted,''
  Jan 2023. [Online]. Available:
  \url{https://www.bloomberg.com/news/articles/2023-01-18/cyber-attack-hits-1-000-merchant-ships-as-norway-firm-targeted?leadSource=uverify+wall}
\BIBentrySTDinterwordspacing

\bibitem{arghire_2023}
\BIBentryALTinterwordspacing
I.~Arghire, ``Ransomware gang publishes data allegedly stolen from maritime
  firm royal dirkzwager,'' Mar 2023. [Online]. Available:
  \url{https://www.securityweek.com/ransomware-gang-publishes-data-allegedly-stolen-from-maritime-firm-royal-dirkzwager/}
\BIBentrySTDinterwordspacing

\bibitem{nguyen2021advanced}
B.~L. Nguyen, T.~Vu, C.~Ogilvie, H.~Ravindra, M.~Stanovich, K.~Schoder,
  M.~Steurer, C.~Konstantinou, H.~Ginn, and C.~Schegan, ``Advanced load
  shedding for integrated power and energy systems,'' in \emph{2021 IEEE
  Electric Ship Technologies Symposium (ESTS)}.\hskip 1em plus 0.5em minus
  0.4em\relax IEEE, 2021, pp. 1--6.

\bibitem{batmani2022sdre}
Y.~Batmani, Y.~Khayat, A.~Salimi, H.~Bevrani, S.~Mirsaeidi, and
  C.~Konstantinou, ``Sdre-based primary control of dc microgrids equipped by a
  fault detection/isolation mechanism,'' \emph{Energy Reports}, vol.~8, pp.
  8215--8224, 2022.

\bibitem{dav}
V.~{Nasirian}, S.~{Moayedi}, A.~{Davoudi}, and F.~L. {Lewis}, ``Distributed
  cooperative control of {DC} microgrids,'' \emph{IEEE Transactions on Power
  Electronics}, vol.~30, no.~4, pp. 2288--2303, 2015.

\bibitem{takiddin2022data}
A.~Takiddin, S.~Rath, M.~Ismail, and S.~Sahoo, ``Data-driven detection of
  stealth cyber-attacks in dc microgrids,'' \emph{IEEE Systems Journal},
  vol.~16, no.~4, pp. 6097--6106, 2022.

\bibitem{sync}
M.~Zhu and S.~Mart{\'\i}nez, ``Discrete-time dynamic average consensus,''
  \emph{Automatica}, vol.~46, no.~2, pp. 322--329, 2010.

\bibitem{sahoo2017distributed}
S.~Sahoo and S.~Mishra, ``A distributed finite-time secondary average voltage
  regulation and current sharing controller for dc microgrids,'' \emph{IEEE
  Transactions on Smart Grid}, vol.~10, no.~1, pp. 282--292, 2017.

\bibitem{song2020recursive}
Z.~Song, A.~Skuric, and K.~Ji, ``A recursive watermark method for hard
  real-time industrial control system cyber-resilience enhancement,''
  \emph{IEEE Transactions on Automation Science and Engineering}, vol.~17,
  no.~2, pp. 1030--1043, 2020.

\bibitem{kiperberg2015remote}
M.~Kiperberg, A.~Resh, and N.~J. Zaidenberg, ``Remote attestation of software
  and execution-environment in modern machines,'' in \emph{2015 IEEE 2nd Int'l
  Conference on Cyber Security and Cloud Computing}.\hskip 1em plus 0.5em minus
  0.4em\relax IEEE, 2015, pp. 335--341.

\bibitem{seshadri2005pioneer}
A.~Seshadri, M.~Luk, E.~Shi, A.~Perrig, L.~Van~Doorn, and P.~Khosla, ``Pioneer:
  verifying code integrity and enforcing untampered code execution on legacy
  systems,'' in \emph{Proceedings of the twentieth ACM symposium on Operating
  systems principles}, 2005, pp. 1--16.

\bibitem{wang2015confirm}
X.~Wang, C.~Konstantinou, M.~Maniatakos, and R.~Karri, ``Confirm: Detecting
  firmware modifications in embedded systems using hardware performance
  counters,'' in \emph{2015 IEEE/ACM International Conference on Computer-Aided
  Design (ICCAD)}.\hskip 1em plus 0.5em minus 0.4em\relax IEEE, 2015, pp.
  544--551.

\bibitem{wen2010implicit}
Y.~Wen, M.~Huang, J.~Zhao, and X.~Kuang, ``Implicit detection of stealth
  software with a local-booted virtual machine,'' in \emph{The 3rd Int'l
  Conference on Information Sciences and Interaction Sciences}.\hskip 1em plus
  0.5em minus 0.4em\relax IEEE, 2010, pp. 152--157.

\bibitem{seol2015trusted}
J.~Seol, S.~Jin, D.~Lee, J.~Huh, and S.~Maeng, ``A trusted iaas environment
  with hardware security module,'' \emph{IEEE Transactions on Services
  Computing}, vol.~9, no.~3, pp. 343--356, 2015.

\bibitem{mclaughlin2016cybersecurity}
S.~McLaughlin, C.~Konstantinou, X.~Wang, L.~Davi, A.-R. Sadeghi, M.~Maniatakos,
  and R.~Karri, ``The cybersecurity landscape in industrial control systems,''
  \emph{Proceedings of the IEEE}, vol. 104, no.~5, pp. 1039--1057, 2016.

\bibitem{rath2020cyber}
S.~Rath, D.~Pal, P.~S. Sharma, and B.~K. Panigrahi, ``A cyber-secure
  distributed control architecture for autonomous ac microgrid,'' \emph{IEEE
  Systems Journal}, vol.~15, no.~3, pp. 3324--3335, 2020.

\bibitem{feng2018behaviorki}
X.~Feng, Q.~Yang, L.~Shi, and Q.~Wang, ``Behaviorki: behavior pattern based
  runtime integrity checking for operating system kernel,'' in \emph{2018 IEEE
  Int'l Conference on Software Quality, Reliability and Security (QRS)}.\hskip
  1em plus 0.5em minus 0.4em\relax IEEE, 2018, pp. 13--24.

\bibitem{ali2020novel}
A.~Ali and M.~M. Yousaf, ``Novel three-tier intrusion detection and prevention
  system in software defined network,'' \emph{IEEE Access}, vol.~8, pp.
  109\,662--109\,676, 2020.

\bibitem{jha2008towards}
S.~Jha, N.~Li, M.~Tripunitara, Q.~Wang, and W.~Winsborough, ``Towards formal
  verification of role-based access control policies,'' \emph{IEEE Transactions
  on Dependable and Secure Computing}, vol.~5, no.~4, pp. 242--255, 2008.

\bibitem{spitzner2003honeypots}
L.~Spitzner, ``Honeypots: Catching the insider threat,'' in \emph{19th Annual
  Computer Security Applications Conference, 2003. Proceedings.}\hskip 1em plus
  0.5em minus 0.4em\relax IEEE, 2003, pp. 170--179.

\bibitem{lakshminarayana2019moving}
S.~Lakshminarayana, E.~V. Belmega, and H.~V. Poor, ``Moving-target defense for
  detecting coordinated cyber-physical attacks in power grids,'' in \emph{2019
  IEEE International Conference on Communications, Control, and Computing
  Technologies for Smart Grids (SmartGridComm)}.\hskip 1em plus 0.5em minus
  0.4em\relax IEEE, 2019, pp. 1--7.

\bibitem{pugliaresi2013us}
L.~Pugliaresi, ``Us house of representatives committee on oversight and
  government reform,'' 2013.

\bibitem{tam2018maritime}
K.~Tam and K.~D. Jones, ``Maritime cybersecurity policy: the scope and impact
  of evolving technology on international shipping,'' \emph{Journal of Cyber
  Policy}, vol.~3, no.~2, pp. 147--164, 2018.

\bibitem{ravindra2016documentation}
H.~Ravindra, M.~Stanovich, and M.~Steurer, ``{Documentation for a notional two
  zone MVDC shipboard power system model implemented on the RTDS},'' 2016.

\bibitem{xie2017comparative}
R.~Xie, R.~Mo, Y.~Shi, and H.~Li, ``Comparative study of the dc-dc power
  conversion module based on dual active bridge converter and modular
  multilevel converter for shipboard mvdc system,'' in \emph{2017 IEEE Electric
  Ship Technologies Symposium (ESTS)}.\hskip 1em plus 0.5em minus 0.4em\relax
  IEEE, 2017, pp. 36--43.

\end{thebibliography}

\end{document}